\documentclass[twocolumn,prd,nofootinbib,showpacs,floatfix]{revtex4}
\usepackage{epsfig}

\newcommand{\be}{\begin{equation}}
\newcommand{\ee}{\end{equation}}
\newcommand{\bea}{\begin{eqnarray}}
\newcommand{\eea}{\end{eqnarray}}

\usepackage[dvips]{color}
\definecolor{Black}{named}{Black}
\definecolor{Red}{named}{Red}

\begin{document}

\title{Oscillations of solar atmosphere neutrinos}

\author{	G.~L.~Fogli$^{1}$, 
		E.~Lisi$^{1}$, 
		A.~Mirizzi$^{1}$, 
		D.~Montanino$^{2}$ and 
		P.~D.~Serpico$^{3}$ 
\medskip\smallskip } 
\affiliation{ $^1$Dipartimento di Fisica and Sezione INFN di Bari,
                Via Amendola 173, 70126 Bari, Italy 
\smallskip \\
			 $^2$Dipartimento di Fisica and Sezione INFN di Lecce, 
			 Via Arnesano, 73100 Lecce, Italy 
\smallskip \\
			$^3$Max-Planck-Institut f\"ur Physik (Werner-Heisenberg-Institut),
			F\"ohringer Ring 6, 80805 Munich, Germany}


\begin{abstract}
The Sun is a source of high energy neutrinos ($E \gtrsim 10$~GeV) produced
by cosmic ray interactions in the solar atmosphere. We study the
impact of three-flavor oscillations (in vacuum and in matter)
on solar atmosphere neutrinos, and calculate their observable fluxes at  Earth,
as well as their event rates in a kilometer-scale detector in water or ice.
We find that peculiar three-flavor oscillation effects in matter, which 
can occur in the energy range probed by solar atmosphere neutrinos, are
significantly suppressed by averaging over the production region and
over the neutrino and antineutrino components. In particular, we
find that the relation between the neutrino fluxes at the Sun and at 
the Earth can be approximately expressed in terms of phase-averaged 
``vacuum'' oscillations, dominated by a single mixing parameter (the angle
$\theta_{23}$).
\end{abstract}
\pacs{ 14.60.Pq,              
       96.50.Sb,              
       95.55.Vj,
       26.65.+t
\hfill Preprint MPP-2006-92}

\maketitle

\newcommand{\nuornubar}{{\stackrel{{}_{(-)}}{\nu}\!\!}}
\newcommand{\nubar}{\bar \nu}
\newcommand{\nui}{\nu_i}
\newcommand{\nuj}{\nu_j}
\newcommand{\nubari}{\nubar_i}
\newcommand{\nubarj}{\nubar_j}
\newcommand{\san}{{\rm SA\nu}}

\section{Introduction}

The Sun is a well-known source of low-energy electron neutrinos
($E\lesssim 20$ MeV), continually produced
in  its core by nuclear reactions. The study of this
flux has 
been one of the most exciting enterprises in astroparticle physics, 
and has provided us with crucial evidence for $\nu$ flavor oscillations 
(see, e.g., \cite{Ba89,Fogli:2006fu} and references therein).

High-energy neutrinos and antineutrinos (with $E\gtrsim 10$~GeV) are also
continually produced by cosmic ray interactions in the solar
atmosphere~\cite{SSG91,Moskalenko:1993ke,Ingelman:1996mj,
Hettlage:1999zr,Learned:2000sw}, which lead to the production
of secondary mesons and their subsequent decay into $\nu_e$,
$\overline\nu_e$, $\nu_\mu$, and $\overline\nu_\mu$, collectively
referred to as ``solar atmosphere neutrinos'' (SA$\nu$) hereafter. 
From pion production dominance,  at the source the SA$\nu$ flavor ratio is 
expected to be $\phi_e:\phi_{\mu}:\phi_{\tau}\simeq 1:2:0$, 
with approximately equal fluxes of $\nu$ and $\overline\nu$. 
Concerning the $\san$ energy spectrum, the Monte Carlo simulation in
\cite{Ingelman:1996mj} finds a power-law decrease
($\sim E^{-3}$) for $E\gtrsim 100$~GeV. At lower energies, heliomagnetic effects
are expected to make the spectrum less steep (as $\sim E^{-2}$) 
\cite{SSG91,Hettlage:1999zr}. 

At the Earth, within the solid angle subtended by the Sun,
the $\san$ flux significantly exceeds the more familiar
``atmospheric neutrino'' flux generated by cosmic rays interacting 
with the Earth atmosphere. In fact,
because of the different scale heights, the density
at the first interaction point is lower in the solar atmosphere
than in the terrestrial one, and therefore a larger fraction of 
mesons can decay into neutrinos instead of being absorbed. 
The expected $\san$ signal can be quantified in terms of 
$O(10)$ events per year in a km$^3$ of water (or ice) above
a threshold $E\gtrsim 100$~GeV~\cite{Ingelman:1996mj}, and might thus
be detected by large-volume $\nu$ telescopes like
IceCube~\cite{Halzen:2006mq}.

The characterization of the $\san$ signal
is an important goal, both in itself and  because it
might represent a ``background'' in searches for signatures 
of possible weakly interacting massive particle (WIMP) annihilation
in the Sun~\cite{Srednicki:1986vj,
 Ellis:1992df,Bergstrom:1998xh,deGouvea:2000un,Crotty:2002mv,Cirelli:2005gh,Halzen:2005ar}. 
 (Predicted WIMP event rates are rather 
 model-dependent, ranging from $10^{-12}$ to
$10^3$ events/year/km$^3$ \protect\cite{Halzen:2005ar}.)

Neutrino oscillations can significantly affect the $\san$ flux, and
must be taken into account in the calculation of the corresponding
lepton event rates in detectors at the Earth.
Earlier studies have stressed, e.g., the relevance of
$\nu_\mu\to\nu_\tau$ vacuum oscillations \cite{Hettlage:1999zr}, which
might produce a few $\tau$ events  per year in a km$^3$ detector
(for $E>100$~GeV), comparable to the $\tau$ event rate in OPERA, the
CERN-to-Gran Sasso long-baseline
experiment~\cite{Sirignano}. Although very challenging from
an experimental viewpoint, 
the identification of $\san$-induced 
$\tau$ events in future $\nu$ telescopes might thus supplement 
laboratory tests of $\nu_\mu\to\nu_\tau$ appearance with their
astrophysical counterpart. Given the potential importance of the
$\san$ signal detection, we think  useful to revisit the impact
of flavor transitions (in vacuum and in matter) on solar atmosphere
neutrinos, in the light of the most recent advances in $\nu$
oscillation physics. In particular, taking into account updated
values for the mass-mixing $\nu$ parameters, we investigate
the three-flavor matter effects associated to $\san$ propagation 
in the Sun.

The plan of our work is as follows. 
In Sec.~\ref{nufluxes} we describe the
input $\san$ fluxes at the Sun, basically taken from Refs.~\cite{Ingelman:1996mj,Hettlage:1999zr}. 
In Sec.~\ref{nuprob} we study 
$\san$ oscillations in vacuum and matter within the standard $3\nu$
mixing framework, and then calculate
their flavor transition probabilities at  Earth by solving numerically the
flavor evolution equations. In Sec.~\ref{detection} we evaluate the
impact of flavor oscillations on the observable events rates in a
km$^3$ neutrino telescope. We summarize our results
in Section~\ref{conclusions}.

In a nutshell, we find that interesting,
genuine three-flavor transitions in matter can take place during the 
$\san$ propagation, but also that their impact is significantly
suppressed by a ``conspiracy'' of various effects, including
spatial and energy smearing, and $\nu/\overline\nu$ 
indistinguishability. In practice,
the final oscillation effects 
can be approximated by (phase-averaged) vacuum flavor transitions,
dominated by the mixing angle $\theta_{23}$.

\section{Solar atmosphere neutrino fluxes}\label{nufluxes}

Cosmic rays hitting the solar atmosphere produce secondary particles
(mainly $\pi$'s, plus $\mu$'s and $K$'s), which then
produce both electron and muon
neutrinos and anti-neutrinos%
\footnote{In the following, we  shall sometimes loosely use the symbol
``$\nu$'' to indicate both neutrinos and antineutrinos. When needed,
a  distinction between $\nu$ and $\overline{\nu}$ is made
to avoid ambiguities.} 
 through their weak decay channels. 
The most recent calculation of the $\san$ flux has been
performed in~\cite{Ingelman:1996mj} by using the Lund Monte Carlo programs
PYTHIA and JETSET for high-energy particle interactions
\cite{Sjostrand:1995iq}.

The fluxes at the source are approximately in the ratio
\begin{equation}
\phi_{e}:\phi_{\mu}:\phi_{\tau}\simeq 1:2:0 \,\ ,
\end{equation}
 as
expected by pion production dominance.
Moreover, an almost equal number of
$\nu$ and ${\overline \nu}$ is expected \cite{Ingelman:1996mj},
\begin{equation}
\nu :\overline{\nu} \simeq 1:1 \,\ .
\end{equation}

After production, $\san$'s are affected by the solar medium
through both absorption on nucleons and oscillations in matter. Absorption 
effects, which play a role only at
energies $E\gtrsim 100$~GeV, depend on the impact parameter $b$ of 
the neutrino trajectory,
\begin{equation}
0 \leq b \leq R_{\odot} \,\ ,
\end{equation}
where $b=0$ corresponds to the Sun center, and $b=R_\odot (=
6.96 \times 10^{5}$~km) to the solar disk border. In particular,
for $E\sim 10^2$--$10^3$~GeV, a flux attenuation of 
$\mathcal{O}(10\%)$ or less is
expected only for cases
with $b\sim 0$ (which have little geometrical weight, since $>50\%$ 
of the flux comes from $b>0.7\,R_\odot$),
while for $E\gtrsim 10^3$~GeV 
absorption effects are sizable at any $b$
\cite{Ingelman:1996mj}. For
$E\gtrsim 100$~GeV, the approximate flavor-independence of the
inelastic neutrino cross sections \cite{Crotty:2002mv,cross} allows
to factorize neutrino absorption from neutrino  
oscillation effects (which are, in principle, entangled) 
\cite{Naumov:2001ci}. Therefore, in the energy range of interest
for our work, absorption effects
can be taken into account through an overall attenuation function
 multiplying the original $\nu$ fluxes \cite{Ingelman:1996mj}, while
 oscillation effects can be applied afterwards. 

We take the absorption-corrected, unoscillated 
fluxes of $\nu_e + \overline{\nu}_e$ and
$\nu_{\mu}+ \overline{\nu}_{\mu}$ from the parametrization in
Eq.~(15) of Ref.~\cite{Ingelman:1996mj}, which is
integrated over the solar disk and is valid for 
neutrino energies $ E> 100$ GeV.
For $E<100\,$GeV, the only calculation of $\san$ fluxes we are aware of is
reported in~\cite{SSG91}, and depends appreciably on cosmic ray transport
properties in the helio-magnetic fields.
Following~\cite{Hettlage:1999zr}, we assume $\phi
\propto E^{-\gamma}$, where $\gamma$ is allowed to vary in 
the range $1.75 < \gamma < 2.45$
to parametrize helio-magnetic uncertainties.

\begin{figure}[t]
\vspace*{-5mm}
\centering
\epsfig{figure=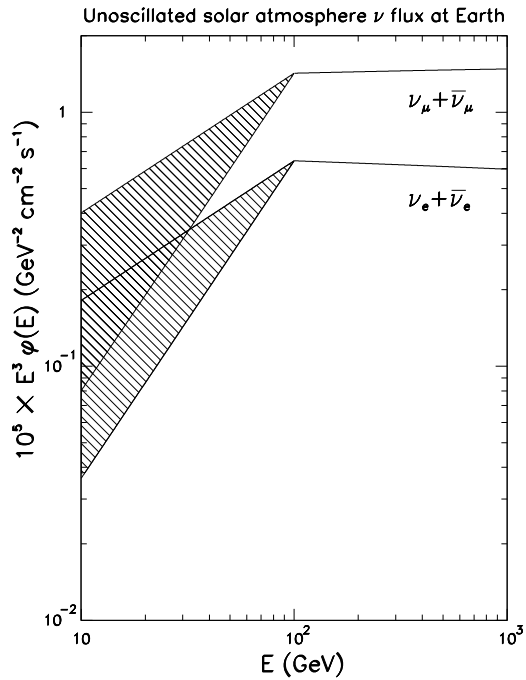,width =0.79\columnwidth,angle=0}
 \caption{\label{fig1} Unoscillated  fluxes of solar atmosphere neutrinos at 
the Earth in terms of neutrino energy. The fluxes are integrated over the Sun
solid angle and summed over the $\nu$ and $\overline\nu$ components (basically equal).
The dominant $E^{-3}$ dependence is factorized out.
 For $E > 100$~GeV, the fluxes are taken from \cite{Ingelman:1996mj}. 
 For $E < 100$~GeV, we take $\phi \propto E^{-\gamma}$ as in~\cite{Hettlage:1999zr}, 
 with $1.75 <\gamma <2.45$ (hatched areas). The overall normalization
 uncertainty (about $20\%$) is not shown.}
\end{figure}

\begin{figure}[t]
\centering
\epsfig{figure=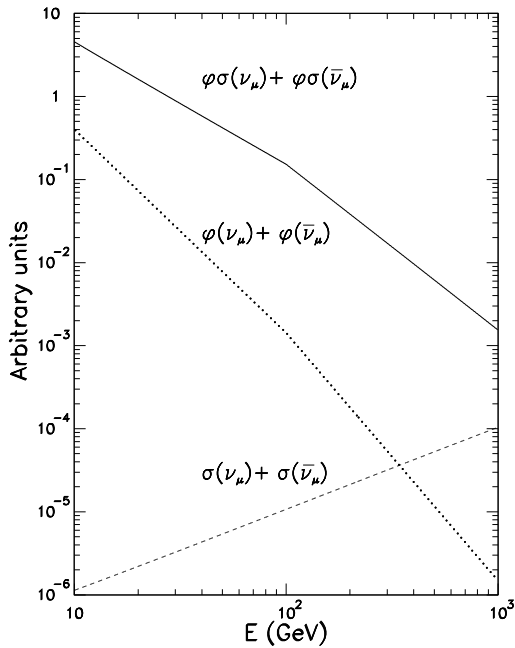,width =0.78\columnwidth,angle=0}
 \caption{\label{fig2} Energy dependence of unoscillated
 $\san$ fluxes (dotted line), of
  charged current interaction cross sections (dashed line), and of their
  product $\phi\, \sigma$ (thick solid line) for 
  solar atmosphere $\nu_{\mu} +\overline{\nu}_{\mu}$, in arbitrary units. Curves
   for $\nu_{e} +\overline{\nu}_{e}$ (not shown) are very similar.} 
\end{figure}

In Figure~\ref{fig1}
we show the unoscillated solar atmosphere $\nu$ fluxes at the Earth,
based on the previous input choices. A typical $\sim 20\%$ 
overall normalization error (not shown), due to primary cosmic ray 
flux uncertainties, should be added to the $\san$ fluxes.
The rapid decrease of the $\san$ fluxes with energy is only partially
compensated by the increase of the neutrino cross sections. 
Figure~\ref{fig2} shows the $\nu_{\mu} + \overline{\nu}_{\mu}$  flux
(with an intermediate spectral index $\gamma= 2.1$ for $E<100$~GeV)
as well as the relevant charged-current interaction cross sections with
the nucleons~\cite{cross}. In the range $10 < E <
10^3$~GeV, the product $\phi\times\sigma$ decreases by more than three orders of
magnitude, and thus the neutrino
event rates above 1 TeV can be ignored in practice.
For the same reason, 
relatively low experimental thresholds will be crucial
for future $\san$ detection.

A final technical comment is in order. We assume a $b$-independent
flavor-ratio at the production, which is strictly true in the limit
in which both pions and muons do not lose energy before 
decaying. We have verified the validity of this approximation for muons
(for pions, given the similar mass but shorter lifetime, the approximation
would be valid {\em a fortiori}). In particular, by using
the solar atmosphere model reported in~\cite{Ingelman:1996mj}, we find
that, in the ``worst'' case ($b\simeq 0$ and $E\simeq 1$~TeV), 
the muon range before decay is at least 20 times smaller
than the typical stopping range for a TeV muon in hydrogen (or
helium)~\cite{Groom:2001kq}. We estimate
that, for $E\lesssim 1\,$TeV,
the isotropic flavor-ratio limit is accurate to better than a few
percent, in agreement with the results of
Ref.~\cite{Ingelman:1996mj}.

\section{Solar atmosphere neutrino oscillations}\label{nuprob}

Solar atmosphere neutrinos are affected
by flavor oscillations (both in matter and in vacuum)
during their propagation. Within the standard  $3\nu$
framework, the oscillation parameters include two squared
mass differences $(\delta m^2,\,\Delta m^2)$ and three
mixing angles $(\theta_{12},\theta_{23},\theta_{13})$, whose
ordered rotations form a unitary mixing matrix $U$.
We take their best-fit values and $2\sigma$ ranges from 
\cite{Fogli2006}:
\begin{eqnarray}
\delta m^2 &=& 7.92\, (1\pm 0.09)\times 10^{-5}\mathrm{\ eV}^2\ ,\\
\Delta m^2 &=& 2.6\,(1^{+0.14}_{-0.15})
\times 10^{-3}\mathrm{\ eV}^2\ ,\\
\sin^2\theta_{12} &=& 0.314 \,(1^{+0.18}_{-0.15})\ ,\label{12}\\
\sin^2\theta_{23} &=&0.45\,(1^{+0.35}_{-0.20})\ ,\label{23}\\
\sin^2\theta_{13} &=& (0.8^{+2.3}_{-0.8})\times 10^{-2}\label{13}\ .
\end{eqnarray}
The parameter $\Delta m^2$ has a physical sign,
corresponding to the
cases of  normal hierarchy
($+\Delta m^2$) or inverted hierarchy ($-\Delta m^2$).
In the following, unless otherwise specified, we assume normal
mass hierarchy, and also set a
possible CP-violating phase $\delta_{\rm{CP}}$ to zero.
Concerning the mixing parameter $\sin^2 \theta_{13}$, we shall
use representative values below the current upper limits. 

\begin{figure}[t]
\centering
\epsfig{figure= 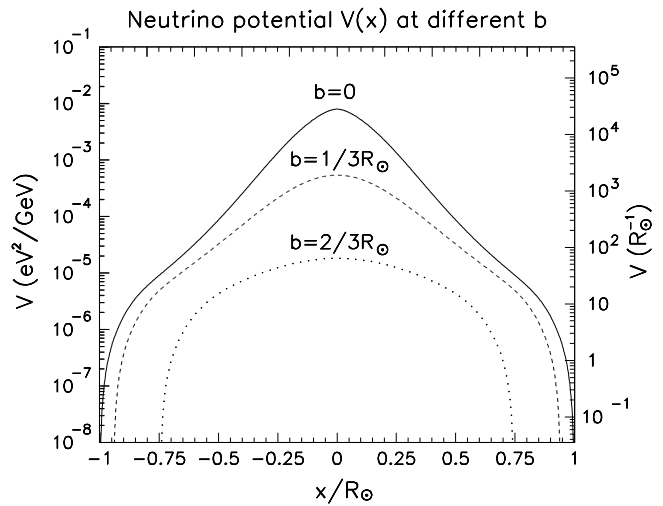,width =0.99\columnwidth,angle=0}
 \caption{Neutrino potential $V(x)$ in the Sun for three
 values of the impact parameter $b$. The $\nu$ trajectory coordinate $x$ is 
 normalized to the solar radius. The potential $V$ is shown both in units
 of eV$^2$/GeV (left axis) and in units of inverse solar radius (right axis).}
 \label{potential}
\end{figure}

In matter, the $\nu$ oscillation dynamics also depends
on the $\nu_e -\nu_{\mu,\tau}$ interaction energy difference 
$V$~\cite{Matt,KuoPa},
\begin{equation}
\label{V}\pm V(x)= \pm \sqrt{2}\, G_F\, N_e(x)\  \;\ (+/- \;\textrm{for}\;
\nu/\overline{\nu})\ ,
\end{equation}
where $N_e$ is the electron density,
which we take from the standard solar model
in \cite{Bahcall:2004pz}, and $x$ is the neutrino
trajectory coordinate in the range $x\in [-x_{\rm{max}},\, +x_{\rm{max}}]$, 
with $x_{\rm{\max}}=(R_{\odot}^2-b^2)^{1/2}$.
The effective potential $V$ encountered by solar atmosphere neutrinos 
depends on the impact parameter $0 \leq  b \leq R_{\odot}$. Figure~\ref{potential} shows 
representative profiles of $V(x)$. Notice that the maximum of $V$ decreases by almost 
three orders of magnitude as $b$ increases from 
0 to $2/3 R_\odot$.

\subsection{The ``vacuum''  approximation}

Let us provisionally ignore matter effects by setting $V(x)=0$ in the Sun
(``vacuum'' approximation, as studied in \cite{Hettlage:1999zr}). 
The oscillated SA$\nu$ fluxes at the detector $D$ are given by
\begin{eqnarray}
\phi^D_e &=& \phi_e P_{ee} + \phi_{\mu} P_{\mu e} \,\ , \\
\phi^D_{\mu} &=& \phi_e P_{e \mu} + \phi_{\mu} P_{\mu \mu} \,\ , \\
\phi^D_{\tau} &=& \phi_e P_{e \tau} + \phi_{\mu} P_{\mu \tau} \,\ ,
\end{eqnarray}
where $P_{\alpha \beta} = P(\nu_{\alpha} \to {\nu_{\beta}})$. 
The two independent vacuum neutrino
oscillations wavelengths, given by $\lambda_L = 4 \pi E/ \delta m^2$ and
$\lambda_H = 4 \pi E/ \Delta m^2$, read
\begin{eqnarray}
\frac{\lambda_L}{R_{\odot}} &=& 4.51 \left(\frac{E}{100 \,\ \textrm{GeV}} \right)
\left( \frac{7.9 \times 10^{-5} \,\ \textrm{eV}^2}{\delta m^2} \right) \ , \\
\frac{\lambda_H}{R_{\odot}} &=& 0.14  \left(\frac{E}{100 \,\ \textrm{GeV}} \right)
\left( \frac{2.6 \times 10^{-3} \,\ \textrm{eV}^2}{\Delta m^2} \right) \ .~~~~~ 
\end{eqnarray}
These wavelenghts are typically much smaller than the Sun-Earth distance
($L \simeq 215\, R_{\odot}$); indeed, in the worst case ($E\simeq 1$~TeV)
one has
\begin{eqnarray}
{L}/{\lambda_L} & \simeq & 5  \,\ , \\
{L}/{\lambda_H} &\simeq & 150  \,\ .
\end{eqnarray}
The occurrence of many oscillation cycles in vacuum ($L/\lambda_{L,H}\gg 1$), 
smeared out by realistic energy resolution effects
($\Delta E/E \gtrsim  10\% $ is expected for km$^3$-sized detectors
\cite{Beacom:2003nh}), guarantees the incoherence of the $\san$ beam at the
Earth (also when matter effects in the Sun are included, as done in the next
Sections).

The vacuum oscillation probabilities are then given by their phase-averaged
expressions,
\begin{equation}
P_{\alpha \beta}^\mathrm{vac} = \sum_{i=1}^3  |U_{\alpha i}|^2 |U_{i \beta}|^2 \,\ .
\end{equation}
If we further take $\sin^2\theta_{13}\simeq 0$ (a reasonable 
approximation in the
context of vacuum oscillations), then 
\begin{eqnarray}
P_{ee}^\mathrm{vac} & \simeq & 1 - 2 s^2_{12} c^2_{12} \,\ , \label{ee}\\
P_{e \mu}^\mathrm{vac} & \simeq & 2 c^2_{12}s^2_{12}c^2_{23} \,\ , \label{emu} \\
P_{\mu \mu}^\mathrm{vac} & \simeq & c^4_{23}(1-2 s^2_{12} c^2_{12}) + s^4_{23} \,\ ,  \label{mumu}\\
P_{\mu \tau}^\mathrm{vac} & \simeq & 2 c^2_{23}s^2_{23}(1- s^2_{12}c^2_{12}) \,\ ,  \label{mutau}\\
P_{e \tau}^\mathrm{vac} & \simeq & 2 c^2_{12} s^2_{12}s^2_{23} \,\ ,\label{etau}
\end{eqnarray}
where $c_{ij} \equiv \cos \theta_{ij}$ and
$s_{ij} \equiv \sin \theta_{ij}$.
 For an initial flux ratio $\phi_{e}:\phi_{\mu}:
\phi_{\tau} \simeq 1:2:0$, the oscillated fluxes at the detector
read
\begin{eqnarray}
\phi^D_e &\simeq & \phi_e  [1+ 2 c^2_{12}s^2_{12}(2 c^2_{23}-1)] \,\ , \\
\phi^D_\mu &\simeq & \phi_{\mu} [1-2 s^2_{23}c^2_{23} + s^2_{12}c^2_{12}c^2_{23}(1-2c^2_{23})] \,\ ,\\
\phi^D_\tau & \simeq & \phi_{\mu} [2 c^2_{23}s^2_{23} + c^2_{12} s^2_{12}s^2_{23}(1-2 c^2_{23})] \,\ ,
\end{eqnarray}
or, numerically, 
\begin{eqnarray}
\phi^D_e &\simeq &   1.043\,  (^{+0.078}_{-0.138})\,  \phi_e\,\ , \label{eD}\\
\phi^D_\mu &\simeq &  0.493\,  (^{+0.050}_{-0.001})\, \phi_{\mu} \,\ ,\label{muD}\\
\phi^D_\tau & \simeq & 0.485\,  (^{+0.022}_{-0.046})\, \phi_{\mu}
\,\ ,\label{tauD}
\end{eqnarray}
where the above central values are calculated for the best-fit values of
 $s^2_{23}$ and $s^2_{12}$, while the (strongly asymmetric)
errors are induced by the  $2 \sigma$ uncertainties of $s^2_{23}$
[see Eq.~(\ref{23})].  This was recently noticed to be a common 
feature for several astrophysical neutrino 
sources~\cite{Serpico:2005sz,Serpico:2005bs,Kachelriess:2006fi}.
The errors induced by the current uncertainties
on $s^2_{12}$ and on $s^2_{13}$ are instead much smaller and
can be neglected 
(see also Sec.~III~D below).
For the specific case of
2-3 maximal mixing ($s^2_{23}=1/2$), one recovers
a well-known result for the phase-averaged, vacuum
neutrino flavor ratio at detection \cite{Athar:2000yw},
\begin{equation}
\phi^D_{e}:\phi^D_{\mu}:
\phi^D_{\tau} \simeq 1:1:1 \ (\mathrm{max.\ mixing})\,\ .
\end{equation}

\subsection{Matter effects}

In the reference energy range $E\in[10,\,1000]$ GeV, solar atmosphere
neutrinos can actually experience peculiar  three-flavor 
oscillation effects in matter which, in principle, may spoil the
``vacuum'' approximation discussed in the previous section.
Matter effects introduce a  characteristic refraction length
$\lambda_r = 2 \pi/V$, which can be expressed as
\begin{equation}
\frac{\lambda_{r}}{R_{\odot}} = 1.78 \times 10^{-6}
\left( \frac{\textrm{eV}^2/\textrm{GeV}}{V} \right)
 \,\ .
\end{equation}
In terms of $\lambda_r$, the oscillation wavelengths in matter
read
 \begin{eqnarray}
 \tilde{\lambda}_{L} &=& \frac{\lambda_L}
{\sqrt{(\cos 2 \theta_{12} \mp \lambda_L/\lambda_{r})^2 +
 \sin^2 2 \theta_{12}}} \,\ , \\
 \tilde{\lambda}_{H} &=& \frac{\lambda_H}
{\sqrt{(\cos 2 \theta_{13} \mp \lambda_H/\lambda_{r})^2 +
 \sin^2 2 \theta_{13}}} \,\ ,
 \end{eqnarray}
where   the $-$($+$) sign refers to neutrinos (antineutrinos).

Potentially large matter effects are
expected when one of the above denominators is small,
(``resonance condition,'' see~\cite{KuoPa}), i.e., roughly when  
\begin{equation}
\lambda_{L,H} \sim \lambda_{r} \,\ .
\label{eq:resonance}
\end{equation}
 In normal (inverted) hierarchy, the
resonance conditions are realized only in the neutrino (antineutrino)
channel. The fact that
$\delta m^2 \ll \Delta m^2$ also implies that
the conditions $\lambda_{L,H} \sim
\lambda_{r}$ are realized at rather different energies.  The 
dynamics of the $3\nu$ system can then be often approximated
in terms of two
$2\nu$ sub-systems, governed  by $(\lambda_L,
\theta_{12})$ and $(\lambda_H, \theta_{13})$ respectively.

At resonance, the oscillation lengths in
matter are numerically given by
\begin{eqnarray}
\frac{\tilde{\lambda}_L^{\rm{res}}}{R_{\odot}} & \!\simeq\! & 0.97\!
\left(\frac{E}{20 \; \textrm{GeV}} \right)\! \left( \frac{7.9 \times
10^{-5}  }{\delta m^2/\textrm{eV}^2} \right)\!
\left(\frac{0.93}{\sin 2 \theta_{12}}\right)\;, \label{L}\\
\frac{\tilde{\lambda}_H^{\rm{res}}}{R_{\odot}} & \!\simeq\! & 0.68\!
\left(\frac{E}{100 \; \textrm{GeV}} \right)\! \left( \frac{2.6 \times
10^{-3} }{\Delta m^2/\textrm{eV}^2} \right)\! 
\left(\frac{10^{-2}}{\sin^2 \theta_{13}}\right)^{\!\!\frac{1}{2}}\!\!,~~~~~ \label{H}
\end{eqnarray}
the latter equation holding for small $\theta_{13}$.  
Very different effects take place, when a
resonance wavelength $\tilde\lambda^{\rm{res}}$ 
is much smaller, comparable, or much larger than the solar radius
$R_\odot$. For $\tilde\lambda^\mathrm{res}\ll R_\odot$ (low-energy, adiabatic
limit) one expects to recover the vacuum case with small, adiabatic
matter corrections. For $\tilde\lambda^\mathrm{res}\sim R_\odot $,
the solar density profile structure is fully probed, and no particular
approximation or limit is expected to work (also because
the profile is non-monotonic, see Fig.~\ref{potential}); this situation is realized
at $E\sim 20$~GeV (for $\tilde\lambda^{\rm{res}}_L$) or at 
$E\sim O(10^2)$~GeV (for $\tilde\lambda^{\rm{res}}_H$), see Eqs.~(\ref{L},\,\ref{H}). Finally, 
for  $\tilde\lambda^\mathrm{res}\gg R_\odot$, one expects again to
recover the vacuum limit but for a different reason: large wavelenghts  
do not ``resolve'' the solar profile structure (extreme nonadiabatic
limit \cite{KuoPa}). 
In our reference range $E\in[10,\,1000]$ GeV, interesting matter
effects are thus expected to emerge at different energies.

For $\theta_{13}=0$ these effects are simplest 
to describe, since they are controlled
only by $(\lambda_L,\theta_{12})$, and just one resonance
condition ($ \lambda_L\sim \lambda_r$) is effective. 
At low energy ($E\lesssim 10$ GeV), it is
$\tilde{\lambda}_L^{\rm{res}} \ll R_{\odot}$, and 
neutrinos oscillate many times in the solar matter, the
phase information being lost. The potential changes slowly over
one wavelength, so that the adiabatic approximation 
(no transition among mass eigenstates in matter) can be applied.
The flavor transition
 probabilities  are then given by
$ P_{\alpha \beta} = \sum_{i}|\tilde{U}_{\alpha i}|^2 |{U}_{i
\beta}|^2 $, where $\tilde{U}$ is the mixing
matrix in matter at the production point. However,
since solar atmosphere neutrinos are produced
in very low density regions, one can take $U\simeq \tilde U$ at the origin,%
\footnote{At typical $\san$ production depths [$O(10^{3})$~km], 
the density of the solar atmosphere is of $O(10^{-6})\,$g~cm$^{-3}$,
and the neutrino potential is negligible for our purposes
[$V\sim O(10^{-10})$~eV$^2$/GeV, not even shown in Fig.~3].
Notice the difference with the more familiar adiabatic limit
for solar neutrinos, where the high density at production (in the   
solar core) makes $\tilde U\neq  U$.}  
 thus recovering the vacuum limit
$P_{\alpha \beta} \simeq \sum_i  |U_{\alpha i}|^2 |U_{i \beta}|^2 $ at
low energy ($E\lesssim 10$~GeV).
At higher energies (say, $E \gtrsim 50$~GeV), it is instead
$\tilde{\lambda}_{L}^{\rm{res}}\gg R_{\odot}$:
neutrino wavelenghts can not resolve the details of the solar matter density
profile, and the vacuum limit applies again \cite{KuoPa}. 
In conclusion, for $\theta_{13}=0$,
significant matter effects are thus expected to emerge only 
around $E\sim \mathrm{few}\times 10$ GeV.

For $\theta_{13}>0$, 
one expects additional matter effects to occur besides the previous ones,
since a second resonance condition ($ \lambda_H\sim \lambda_r$)
becomes effective. These effects, driven by
$(\lambda_H,\theta_{13})$, are
expected to build up at energies somewhat higher than for 
$\theta_{13}=0$ [e.g., the relevant
energy range is around $E\sim 100$~GeV for $s^2_{13}\simeq 10^{-2}$, 
see Eq.~(\ref{H})]. 
It should be noted that: (1) the two resonances, 
governed by different oscillation parameters,
have
different widths; (2) each is realized twice along the density
profile (with partial cancellation effects); and (3) they may take 
place at energies not widely separated (e.g., $~20$ and
$~100$ GeV), and thus may be partly entangled and
non-factorizable. These complications prevent a straightforward 
generalization or construction of analytical recipes for calculations,
in the spirit of those derived, e.g., in
the context of solar neutrinos~\cite{quasivac}.  
For this reason, in the following section we
shall adopt numerical techniques in order 
to obtain 
quantitative results for the oscillation probabilities
in the solar interior.

Finally, we mention that, in our reference
energy range $E\in [10,\,1000]$~GeV, possible matter effects in the
Earth (before detection) are negligible, except perhaps at the lowest 
energies
($E \simeq 10$--20~GeV, see e.g.\ \cite{Akhmedov:2005zh}), where
they would be anyway much smaller than helio-magnetic flux systematics
(see Fig.~\ref{fig1}). For the
specific case of IceCube, at the South Pole, neutrino trajectories
in the Earth would also be relatively short. In conclusion, 
we can safely ignore Earth matter effects in the following.

\subsection{Neutrino oscillation probabilities}

To evaluate the $\san$ oscillation probabilities
 $P_{\alpha \beta}$,
 we evolve  numerically---through Runge-Kutta routines---the 
 $\nu$ and $\overline\nu$ flavor evolution
equations~\cite{KuoPa}  
for many different values of the impact parameter $b$, corresponding
to different solar matter profiles (see Fig.~\ref{potential}).  More
precisely, we calculate first the amplitudes $A_\odot(\nu_\alpha\to\nu_i)$ 
from the
production point to the Sun exit, where the $\nu_i$'s are the mass eigenstates
in vacuum.
Then we  obtain the 
flavor oscillation probabilities
$P_{\alpha \beta}$ at the Earth as
\begin{equation}
P_{\alpha \beta} = \sum_{i=1}^3   |A_\odot(\nu_{\alpha} \to
\nu_{i})|^2\cdot|U_{i \beta }|^2 \ .
 \end{equation}
The above factorization of probabilities~\cite{KuoPa} from the production 
point to the Sun exit ($|A_\odot|^2$) and from the Sun exit to the 
detector at the Earth
($|U|^2$) is guaranteed by the suppression of all interference
terms, due to the many oscillation cycles in vacuum and to the finite
energy resolution.

From the unitarity property of the probabilities, one can express all
the  $P_{\alpha \beta}$'s in terms of only four independent
quantities~\cite{deGouvea:2000un}. In the following, we choose 
 $P_{ee}$, $P_{e \mu}$, $P_{\mu \mu}$ and
$P_{\mu \tau}$ as independent set.

\begin{figure}[t]
\centering
\epsfig{figure=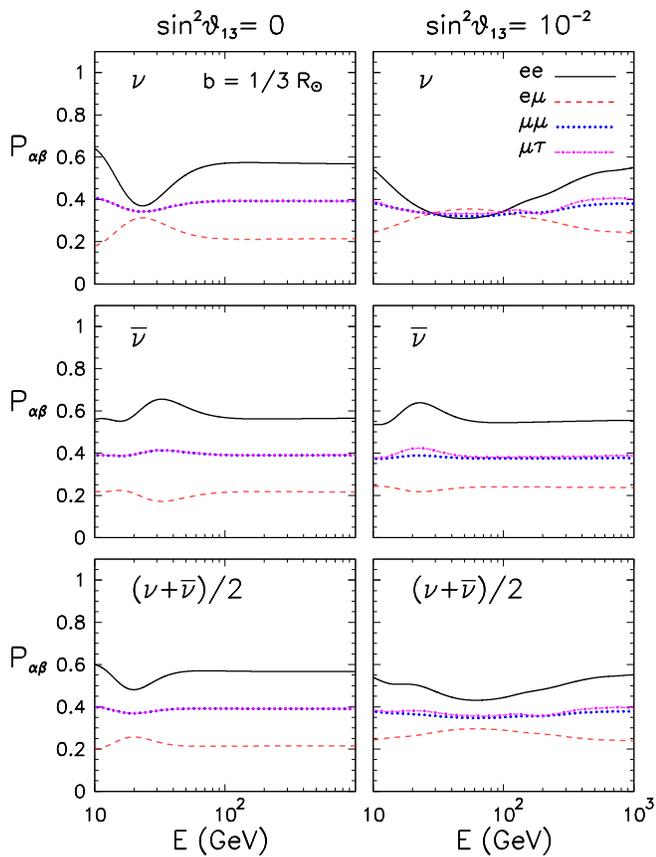,width =0.99\columnwidth,angle=0}
 \caption{\label{fig4} Oscillation  probabilities for solar atmosphere neutrinos
 as functions of energy, for
 $\sin^2 \theta_{13}=0$ (left  panels) and $\sin^2 \theta_{13}= 10^{-2}$
 (right panels). We assume maximal 2-3 mixing, normal hierarchy and
  fix the impact parameter at $b= 1/3 R_{\odot}$. Probabilities are shown 
for $\nu$ (upper panels), $\overline{\nu}$
(middle panels), and their average $(\nu  + \overline{\nu})/2$
 (lower panels).}
\end{figure}

In Figure~\ref{fig4} we plot these
probabilities as functions of neutrino energy $E$
after gaussian energy resolution smearing (with
$\Delta E/E=10\%$ for definiteness),
for a representative impact
parameter $b= 1/3$ $R_{\rm{\odot}}$, and for both 
$s^2_{13}=0$ (left panels) and
 $s^2_{13}=10^{-2}$ 
(right panels), assuming maximal 2-3 mixing and
normal mass hierarchy in both cases.
The upper and middle panels refer to neutrinos and antineutrinos,
respectively. Since solar atmosphere $\nu$'s and $\overline\nu$'s 
have approximately equal fluxes, and are basically indistinguishable
in neutrino telescopes, we find it useful to 
plot the arithmetic average of the
$\nu$ and $\overline\nu$ oscillation probabilities in the lower 
panels of Fig.~\ref{fig4}.

For $\sin^2\theta_{13}= 0$ (left panels of Fig.~\ref{fig4}), the vacuum
limit is reached for $E\gtrsim 50$~GeV, as discussed
in the previous section. The same limit applies at very low
energies ($E\lesssim 10$~GeV, not shown). In the vacuum limit, 
the probabilities
$P_{\alpha\beta}$ are given by Eqs.~(\ref{ee}--\ref{etau}), which
numerically yield:  $P_{ee}\simeq 0.57$,
$P_{e \mu}\simeq 0.22$, $P_{\mu \mu} = P_{\mu \tau} \simeq 0.39$.
Matter effects, as expected, emerge around $E\sim 20$~GeV, and are
more pronounced  in the neutrino channel (where the resonance 
condition $\lambda_L\sim \lambda_r$ can be realized). Notice that, 
at all energies, $P_{\mu\mu}=P_{\mu\tau}$. This is due
to the fact that, for $\sin^2\theta_{13}= 0$, 
$\nu_e \leftrightarrow
\nu_{\mu, \tau}$ oscillations reduce to a  pure two-family case
(driven by $\delta m^2$ and $\theta_{12}$), and the
probabilities in matter can be expressed in terms of $P_{ee} = P_{ee}(\delta
m^2, \theta_{12}, V)$ alone~\cite{Akhmedov:2005zh}. In particular,
after phase averaging, one gets~\cite{Peres:1999yi}
 \begin{eqnarray}
 P_{\mu \mu} &=& s_{23}^4  +P_{ee}c^4_{23}  \,\ , \\
P_{\mu \tau} &=&  (1+P_{ee}) s^2_{23}c^2_{23}
\,\ , \label{eq:plow}
 \end{eqnarray}
and thus
$P_{\mu \mu}= P_{\mu \tau}$ for $s^2_{23}=1/2$.
In addition, one gets $P_{e \mu}=(1-P_{ee})c^2_{23}$~\cite{Peres:1999yi}, and
thus ``opposite'' variations of $P_{e \mu}$ and $P_{ee}$, as evident in the left
panels of Fig.~\ref{fig4}.

For  $\sin^2\theta_{13}= 10^{-2}$ (right panels of Fig.~\ref{fig4}),
additional three-flavor mixing effects, due to a second resonance, are seen to emerge in the
neutrino channel around $E\sim 100$ GeV, with broad features entangled with 
the previous, lower resonance range. As compared 
with the left panels, the vacuum limit is reached at
 higher energies, but with almost the same limiting values
for $P_{\alpha\beta}$. Most of the additional matter effects appear
in $\nu_e$-related oscillations, while $P_{\mu\mu}$ and $P_{\mu\tau}$
change little, the previous equality $P_{\mu\mu}=P_{\mu\tau}$ being
only slightly perturbed.

The two lowest panels in Fig.~\ref{fig4} show that matter
effects, averaged over $\nu$ and
$\overline\nu$ components, produce only mild variations 
with respect to the vacuum limit, independently of the value
of $\sin^2\theta_{13}$. For inverted mass hierarchy, matter
effects would be more prominent for antineutrino oscillation
probabilities,
but their average over $\nu$ and $\overline\nu$ would not
be much different from Fig.~\ref{fig4}. Therefore, even if
we could select a specific solar atmosphere neutrino
impact parameter $b$, the experimental $\nu/\overline\nu$
indistinguishability would make it hard to find signatures
of matter effects, despite the fact that they may be 
potentially large on some $P_{\alpha\beta}$'s. 

Integration over the impact parameter $b$ is expected
to strengthen these conclusions, since trajectories with
relatively high impact parameter $b$ (and thus with short
path in the Sun and small matter effects) have large
geometrical weight. 
Indeed, given the limited angular resolution of neutrino telescopes
($\Delta \theta \gtrsim 1^{\circ}$~\cite{Beacom:2003nh}), the
solar disk angle
($\Delta \theta_{\rm{sun}}\simeq 0.26^{\circ}$) cannot be resolved, and
the neutrino oscillation probabilities must be averaged over $b$:
\begin{equation}
\langle P_{\alpha \beta} (E) \rangle = \frac{1}{ \pi R_{\odot}^2}
\int_{0}^{R_{\odot}} 2 \pi b \; db \; P_{\alpha \beta} (E,b) \,\ .
\end{equation}

\begin{figure}[t]
\centering
\epsfig{figure=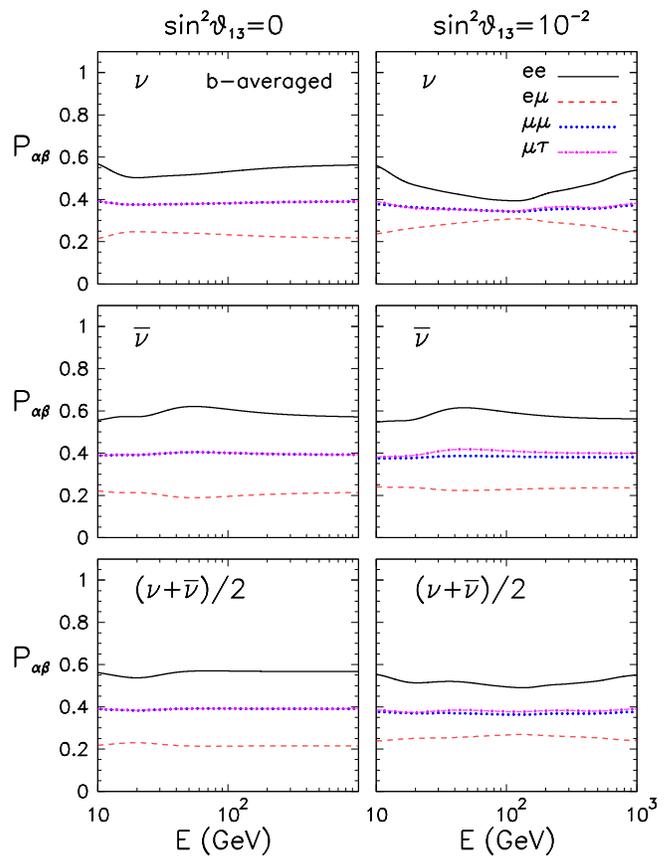,width =0.99\columnwidth,angle=0}
 \caption{\label{fig5} Oscillation  probabilities for solar atmosphere neutrinos,
 averaged over the impact parameter $b$, in
  the same format of Fig.~\ref{fig4}.\medskip\medskip\medskip}
\end{figure}

Figure~5 shows the $b$-averaged oscillation probabilities, in the same
format of Fig.~4. It appears that possible features associated to
matter effects are strongly smeared out in the observable 
$\nu+\overline\nu$ channel. The energy dependence of
all probabilities is rather flat and roughly insensitive
to $\sin^2\theta_{13}$,
with only minor deviations from
the vacuum limit governed by the parameter $s^2_{23}$
[Eqs.~(\ref{ee}--\ref{etau})]. Further considerations
about realistic experimental conditions (integration above 
a given energy threshold, small expected statistics)
can only strengthen such conclusions.


\subsection{Variations of the oscillation parameters}

The previous results show that observable
$\san$ flavor transition effects are dominated by the parameter  
$\sin^2\theta_{23}$. Its relatively large experimental errors 
[Eq.~(\ref{23})] represent, at present, the main 
uncertainty associated to oscillation effects in calculating both
absolute and relative $\san$ fluxes. In order to illustrate
this point we choose a representative quantity which might be of  
interest in the future (should
$\nu$ telescopes be able to discriminate electron events),
namely, the electron-to-muon neutrino flux ratio at the
detector.

Figure~6 shows the $\phi^D_e/\phi_\mu^D$ ratio (with fluxes summed
over their $\nu$ and $\overline\nu$ components) as a function of the neutrino
energy. The three oscillation parameters $(\Delta m^2,\,\delta m^2,\,\sin^2\theta_{12}$) are fixed at their best-fit values, 
and normal hierarchy is assumed. The upper and lower sets of curves  
are obtained for the $\pm2\sigma$ extremal values of $\sin^2\theta_{23}$
from Eq.~(\ref{23}). Within each set of curves, $\sin^2\theta_{13}$
is also taken at its $\pm2\sigma$ extrema from Eq.~(\ref{13}), while
the unknown phase $\delta_\mathrm{CP}$ is set at either $0$ or $\pi$
(for $\sin^2\theta_{13}\neq 0$). The spread of the curves due to the 
$\sin^2\theta_{23}$ uncertainties amounts to about $\pm 15\%$, while
the variations due to $(\sin^2\theta_{23},\delta)$ are roughly a factor three
smaller.
The $\pm 2\sigma$ errors of the other oscillation parameters $(\Delta m^2,\,\delta m^2,\,\sin^2\theta_{12}$), as well as the change from
normal to inverted hierarchy, would
produce even smaller variations in the above curves (not shown). In practice,
one can currently evaluate the effects of oscillation parameter uncertainties 
in terms of $\sin^2\theta_{23}$ only. In the future, when $\sin^2\theta_{23}$
will be determined much more accurately, one might need 
to account also for the (now subleading) errors of
$(\sin^2\theta_{13},\delta_\mathrm{CP})$. In the very far
future, one might even hope that high-statistics samples of $\san$
events could turn these effects from ``uncertainties'' to ``observables.''


\begin{figure}[t]
\centering
\epsfig{figure=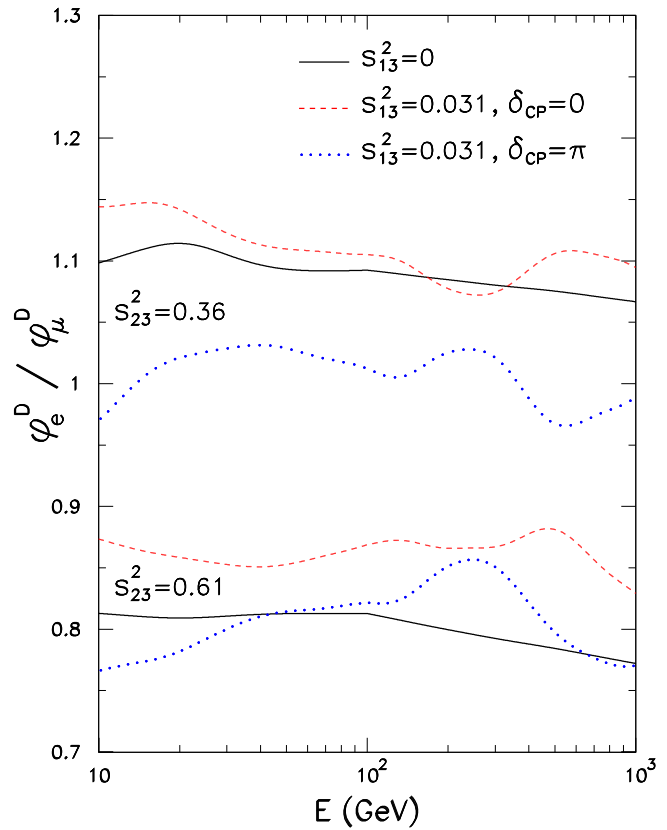,width =0.99\columnwidth,angle=0}
 \caption{\label{fig6} 
Flux ratio $\phi_e^D/\phi_\mu^D$  as a function of 
energy, for normal hierarchy. 
The spread of the curves reflects the $\pm 2\sigma$
spread of $\sin^2\theta_{23}$ and, to a lesser extent,
of $\sin^2\theta_{13}$ and $\delta_{CP}$. 
The spread of the other oscillation parameters 
$\Delta m^2$, $\delta m^2$, and $\sin^2\theta_{12}$ (here fixed to
their best-fit values), as well as a change to inverted hierarchy, 
would produce only minor variations in the above curves (not shown).
\medskip\medskip}
\end{figure}

\subsection{Summary of oscillation effects}

We summarize this Section as follows. We have shown that solar 
atmosphere neutrinos are, in principle, subject to
 significant matter effects in the Sun,
at least in some oscillation channels and for relatively 
central trajectories. These effects can be
sensitive to $\theta_{13}$ and to both squared mass 
differences $\delta m^2$ and $\Delta m^2$, as well as
to the neutrino mass hierarchy and to $\delta_{\rm{CP}}$. However, by averaging over
the neutrino and antineutrino channel, as well as over the
impact parameter, these effects are significantly reduced. 

For approximate estimates and for practical purposes, one can simply adopt the
vacuum limit in Eqs.~(\ref{ee}--\ref{etau}), and take
into account only the uncertainties of $\theta_{23}$
(the spread of the other oscillation parameters being currently
much less relevant). Numerically, one can take: 
\begin{eqnarray}
P_{ee} & \simeq & 0.57  \,\ , \label{qee}\\
P_{e \mu} & \simeq & 0.43\,c^2_{23} \,\ , \label{qemu} \\
P_{\mu \mu} & \simeq & 0.57\,c^4_{23} + s^4_{23} \,\ ,  \label{qmumu}\\
P_{\mu \tau} & \simeq & 1.57\,c^2_{23}s^2_{23} \,\ ,  \label{qmutau}\\
P_{e \tau} & \simeq & 0.43\,s^2_{23} \,\ ,\label{qetau}
\end{eqnarray}
where, currently, $s^2_{23}\in [0.36,\,0.61]$ at $2\sigma$ [Eq.~(\ref{23})].
This amounts, e.g., to approximate the curves in the lower panels
of Fig.~5 with flat, horizontal lines. Of course,
when $\san$ will be detected in future experiments, it will 
be worthwhile to make the best use of the data
by fully including matter effects in the oscillation analysis, in
order to avoid unnecessary approximations. 
For the sake of
accuracy, we do include matter effects in the 
calculations discussed below.

\section{Fluxes and event rates at Earth}\label{detection}

\begin{figure}[t]
\centering
\epsfig{figure=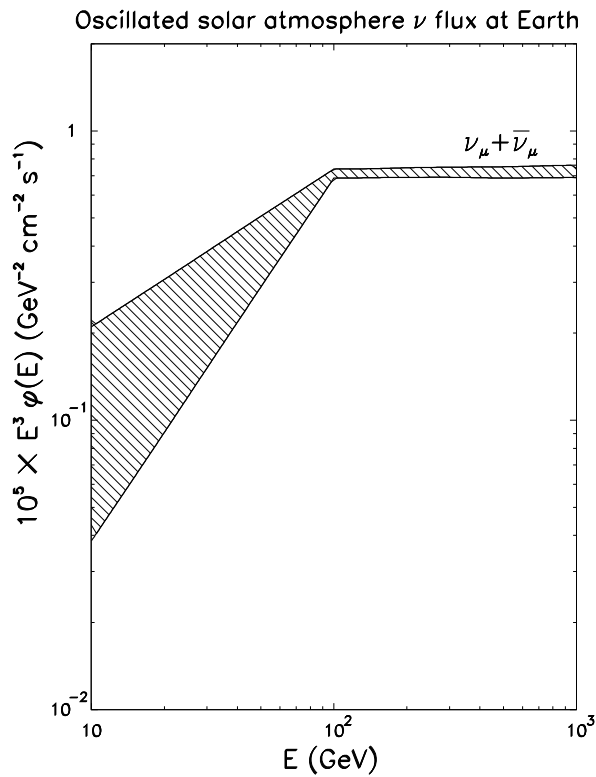,width =0.9\columnwidth,angle=0}
 \caption{\label{fig7} Oscillated $\nu_{\mu} + {\overline{\nu}}_{\mu}$  
 flux of solar atmosphere neutrinos at the Earth,
 integrated over the solid angle of the Sun. The amplitude of the band above 
 100 GeV corresponds to the current $\pm2\sigma$ uncertainties of the
mixing parameter $\theta_{23}$; below 100~GeV, the dominant
uncertainty is instead related to the energy spectral index.
}
\end{figure}

Given the unoscillated input fluxes of Sec.~II and the neutrino oscillation
probabilities discussed in Sec.~III, one can finally obtain the  
$\san$ fluxes and event rates observable at the Earth.
Figure~\ref{fig7} shows the 
oscillated $\nu_{\mu}+\overline{\nu}_{\mu}$ flux at the Earth,
including matter effects (the results with no matter effects
would be, as already mentioned,
 very similar). The amplitude of the band in Fig.~7 
includes the $2\sigma$ uncertainties
in the mixing parameter $\theta_{23}$ from Eq.~(\ref{23}). 
Below $\sim 100$~GeV, it also includes the dominant uncertainty
 related to the energy spectral index~$\gamma$. 
In particular,
the upper curve in Fig.~7 is obtained for $s^2_{23}= 0.61$
and $\gamma=2.45$, while the lower curve is obtained
for $s^2_{23}= 0.36$ and spectral index $\gamma=1.75$. 
The fluxes of $\nu_e+\overline\nu_e$ and $\nu_\tau+\overline\nu_\tau$ (not shown)
would be very similar to the flux
of $\nu_\mu+\overline\nu_\mu$, since the flavor ratio at the detection 
is roughly $1:1:1$ around maximal mixing. To all such fluxes,
one should add at least a $\sim 20\%$ normalization error (not shown)
due to the uncertainties of the primary cosmic 
ray flux, and to the finite 
statistics of the flux simulations in~\cite{Ingelman:1996mj}.

Solar atmosphere neutrinos generate events observable in large-volume
detectors above a certain energy threshold  $E_\mathrm{th}$. The
corresponding event rates (for flavor $\alpha$) are given by
\begin{equation}
R_{\alpha} = \int_{E_{\rm{th}}}^{\infty} dE\; \phi_{\alpha}^D(E)\; \sigma_{\alpha}(E)\;
\frac{\rho}{m_p}\; L_{\alpha}(E)\; A \,\ ,
\end{equation}
where $\rho$ is the density of the medium ($\rho \simeq 1$~g/cm$^3$
in water or ice), $m_p$ is the proton mass 
(so that $\rho/m_p$ is the target nucleon number density), $\sigma_\alpha$ is
the cross section, $L_\alpha(E)$ is the range of the produced lepton
(or the detector thickness $h$, whichever the larger),
and $A$ is the effective area of the detector. The charged-current
 interaction cross sections
$\sigma_\alpha$ are taken from~\cite{cross}. For a
IceCube-like detector we take
$A= 1$~km$^2$ and $h=1$~km. Concerning the lepton range,
we can take $L_e=h=L_\tau$, whereas for $\nu_{\mu}$
\begin{equation}
L_{\mu}(E) = \textrm{max}\left \{\frac{1}{\beta \,\rho}\, \ln
\frac{E + \alpha/\beta}{E_{\rm{th}} + \alpha/\beta} \,,\,\ h\right\} \,\ ,
\end{equation}
where  the first term in brackets
represents the distance travelled by a muon before its energy drops below 
the threshold $E_{\rm{th}}$,  the parameters
 $\alpha=2.5$~MeV$/$(g~cm$^{-2}$) and $\beta= 4.0 \times
10^{-6}$~(g cm$^{-2}$)$^{-1}$  being
adopted from~\cite{Ingelman:1996mj,Hettlage:1999zr}.
More accurate calculations of the event rates would require 
MonteCarlo simulations for specific detector settings, which
are beyond the scope of this work.

\begin{figure}[t]
\centering
\epsfig{figure=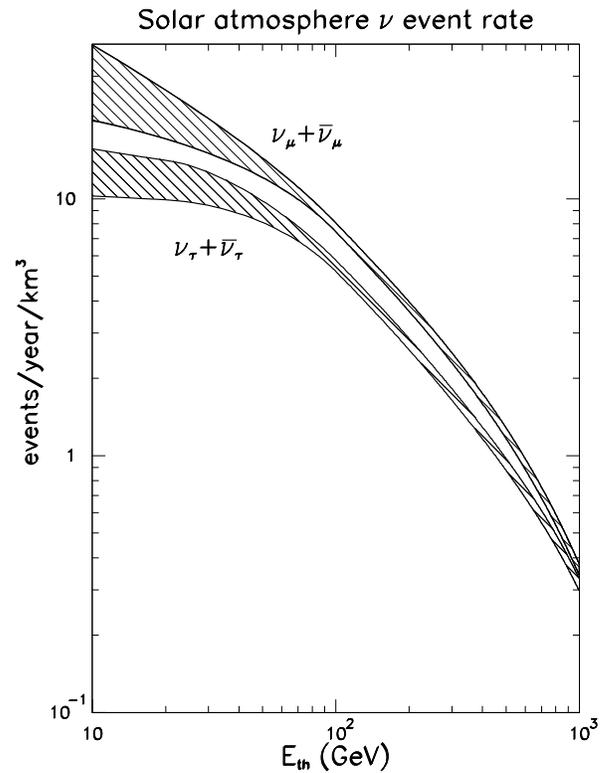,width =0.9\columnwidth,angle=0}
 \caption{\label{fig8} $\nu_{\mu}$ and $\nu_{\tau}$ event rate in a
 km$^3$ detector, as a function of the detector threshold energy $E_{\rm th}$.
 The $\nu_e$ rate (not shown) is very similar to the one of $\nu_{\mu}$.
 The width of each band reflects the uncertainties on the initial fluxes
  and on the mixing parameters. See the text for details.}
\vspace*{-3mm}
\end{figure}

Figure~\ref{fig8} shows the results of our calculation for the
muon and tau lepton event rates (summing over $\nu$ and $\overline\nu$),
as
functions of the threshold energy $E_\mathrm{th}$,
including oscillations in matter. 
(If the vacuum limit were adopted,
the curves in Fig.~8 would change by $<3\%$.)
The rate of electron 
events (not shown) would be very similar to that of muon events. 
The width of the bands in Fig.~8 includes, as in Fig.~7,
the uncertainties associated to $s^2_{23}$ and to $\gamma$.
On top, one should add a systematic normalization uncertainty 
of $\sim \pm 20\%$ (not shown). 
It appears that a relatively low energy threshold 
(say, at or below 100 GeV) is crucial to get a handful of events per year
in km-sized detectors. Note that the rate in Fig.~8 includes events
from all arrival directions,
both above the horizon (``downgoing'') and below the horizon
(``upgoing''). 

We conclude this Section with some qualitative remarks about 
future observations of solar atmosphere neutrino events with rates as
low as those calculated in Fig.~8, which require careful
background rejection.

The atmospheric neutrino background can be partly
rejected by directional cuts (in the Sun's direction). Unfortunately,
the typical lepton scattering angle and detector angular
resolution are larger than the solar disk angle.
Therefore, despite the fact that
the $\san$ signal dominates over the atmospheric neutrino
noise in the solar disk radius, the signal-to-background ratio drops
to $\alt 1$ in realistic angular bins~\cite{Hettlage:1999zr}.
In addition, rejection of the muon background from cosmic rays showers 
(still huge at $10^2$--$10^3$~GeV) requires the selection of 
upgoing events, with corresponding reduction
of the event rates. However, even a few $\san$-induced muon
events per year above 
a threshold of about 100 GeV (versus a prediction twice as large
in the absence of $\theta_{23}$-driven oscillations)
might make it possible, in a decade or so, the ``solar atmospheric
neutrino check'' of the
large 2-3 mixing established by terrestrial atmospheric neutrinos.

When a $\san$ signal will be established,
a rough spectral analysis at different energy thresholds might
also help to check the current expectations for energy spectral index.
In this respect, our work clearly shows that neutrino oscillations 
cannot be responsible for dramatic variations of the
$\san$ rates with energy; if found, such variations should then be ascribed
to other factors (e.g., unexpectedly large helio-magnetic effects above 100 GeV).

The uncertainties on the normalization and energy spectrum
of the $\san$ flux can also be reduced
through solar gamma-ray data, since they have common production
mechanisms (e.g., the $\gamma$'s from $\pi^0$ decays). The upper limit sets 
by EGRET on the gamma flux above 100~MeV
($\Phi_\gamma <2\times 10^{-7}\,$ cm$^{-2}$ s$^{-1}$) \cite{EGRETmoon}
is only a factor $\sim 3$ above the nominal predictions of~\cite{SSG91}.
The galactic diffuse background in the same angular 
bin is about one order of magnitude
smaller, and with the performances expected by the forthcoming GLAST satellite
mission~\cite{GLAST}, the solar gamma signal should be easily detectable. This measurement
may be complemented by the detection of the inverse Compton scattering photons 
produced by cosmic ray electrons, which probe the more external layers of solar 
atmosphere~\cite{Moskalenko:2006ta}. All in all, from gamma-ray astronomy
we expect a relevant boost in the characterization of the solar heliosphere, with 
consequent sharpening of the predictions of $\san$ fluxes.

Finally, let us briefly mention a technical point. Since
the Sun works as a target for cosmic rays (thus generating $\san$'s), 
it also shields the Earth from cosmic rays 
coming from the same solid angle 
(thus depleting atmospheric $\nu$'s in that direction). 
One might then wonder whether, within the solar solid angle,
the sought $\san$ signal is accidentally compensated by the 
atmospheric $\nu$ depletion due to the Sun's ``shadow,''
thus escaping detection.
We argue that this is not the case, for at least two reasons:
$(i)$ the atmospheric $\nu$ background depletion
is almost one order of magnitude smaller than the signal 
one is looking for; $(ii)$ 
interplanetary magnetic fields displace (and smear) the 
Sun shadow.
The exact displacement is difficult to predict, but
the one measured with muons by MACRO~\cite{Ambrosio:2003mz}
amounts to about 0.6$^\circ$ for median energies of the proton primaries
of $\sim 22$ TeV. For the lower energies we are focusing on, we expect
larger displacements, which may allow
to separate the Sun shadow in the atmospheric $\nu$ background
from the position of the $\san$ excess.
Finally, the position of this dip (and the similar one due to
the Moon shadow) will be tagged more accurately in neutrino telescopes
through downgoing muons, and thus the deficit  
should be easily corrected for.


\section{Conclusions and prospects}\label{conclusions}
During the last forty years, the Sun has been the most important source of
astrophysical neutrinos, which have provided us with crucial
evidence of neutrino oscillations. 
Neutrino astrophysics is now entering 
 a new era,  where next generation neutrino telescopes
 will be capable to detect high-energy 
 (extra)galactic astrophysical neutrino fluxes
from new sources. These $\nu$'s would be useful both to
 probe the properties of the sources~\cite{Barenboim} and to test for new physics 
beyond the standard  three-neutrino
 mixing scenario~\cite{decay}.

The same telescopes might be able
to ``see'' the Sun in a new way, namely, as a source of
high energy neutrinos, produced by cosmic ray interactions in the
solar atmosphere ($\san$). Even if only a handful of events per year 
may be detectable in a km$^3$ telescope, their theoretical 
fluxes are relatively under control, and thus a measurement
might provide a high-energy, astrophysical probe of neutrino
flavor mixing.
 Moreover, the $\san$ flux represents an unavoidable background
in searching possible
 neutrino fluxes induced by WIMP annhihilation in the Sun core. 
For these reasons, it appears worthwhile to
 characterize the $\san$ flux detectable at the Earth, after
neutrino oscillations have occurred in the solar matter
and in vacuum.

In our work we have studied solar atmosphere neutrino oscillations in
detail, in the light of the most recent determinations of the
neutrino mass mixing parameters, and paying particular attention
to matter effects (generally neglected a priori in this context).
We have found that, in the energy range $E\in[10,\,1000]$~GeV
relevant for detection, $\san$ oscillations in matter have in
principle a rich phenomenology, and can generate effects 
which depend on all $3\nu$ mass-mixing parameters. Unfortunately,
these effects are smeared out to a large extent by summing over
the neutrino and antineutrino channel, and by integrating over the 
solar disk angle (as well as over energy). 
In practice, oscillations are dominated
by the ``vacuum'' effect of the $\theta_{23}$ mixing angle. 
Therefore, even if matter effects can and do occur,  the
original flavor ratio $\phi_{e}:\phi_{\mu}:\phi_{\tau}\simeq 1:2:0$ at 
the source approximately yields the well-known ratio
$\phi^D_{e}:\phi^D_{\mu}:\phi^D_{\tau}\simeq  1:1:1$ at the
detector (for maximal 2-3 mixing), 
just as in the case
of phase-averaged vacuum oscillations.

Further refinements may include improved estimates
of the input $\san$ fluxes and of their uncertainties.  
In recent years, significant
progress has been made in determining the cosmic 
ray composition and
in refining the interaction and cascade development models~\cite{Stanev:2004ys}. 
More detailed solar atmosphere models 
\cite{Gu97} and
future gamma ray data might also help to constrain the
$\san$ production mechanism. 
In particular, observations in gamma rays may sharpen our knowledge of 
helio-magnetic effects \cite{SSG91,Moskalenko:2006ta}.
Earlier estimates 
of the unoscillated $\san$ fluxes 
\cite{SSG91,Moskalenko:1993ke,Ingelman:1996mj,Hettlage:1999zr}
could thus be usefully revisited. This task will certainly
become mandatory if a $\san$ signal will be found 
in neutrino telescopes or, even better, if it will become
a background for signatures of WIMP annihilation in the Sun.

\medskip
\begin{acknowledgments}
We thank M.\ Maltoni and G.\ Sigl for useful comments. 
This work is supported in part by
the Italian ``Istituto Nazionale di Fisica Nucleare'' (INFN) and by the ``Ministero dell'Istruzione,
Universit\`a e Ricerca'' (MIUR) through the ``Astroparticle Physics'' research project.
The work of P.D.S.\ is supported  in part by the Deutsche Forschungsgemeinschaft
under grant No.~SFB 375 and by the European Union under the ILIAS
project, contract No.~RII3-CT-2004-506222. 
A.M.\ acknowledges kind hospitality at the 
 Max-Planck-Institut f\"ur Physik (Munich, Germany), where this work was
 completed.
\end{acknowledgments}



\begin{thebibliography}{00}


\bibitem{Ba89}  J.N.~Bahcall, {\em Neutrino Astrophysics\/}
                (Cambridge U.\ Press, Cambridge, England, 1989).


\bibitem{Fogli:2006fu}
  G.~L.~Fogli, E.~Lisi, A.~Marrone and A.~Palazzo,
  ``Solar neutrinos (with a tribute to John N.\ Bahcall),''
  Proceedings of {\em NO-VE~2006\/}, 3rd International Workshop
  on Neutrino Oscillations in Venice (Venice, Italy, 2006),
  ed.\ by Milla Baldo Ceolin (Edizioni Papergraf, Padova, Italy, 2006),
  p.~69  [hep-ph/0605186].



\bibitem{SSG91}
D.~Seckel, T.~Stanev, and T.~K.~Gaisser, ``Signatures of cosmic-ray
interactions on the solar surface,'' Astrophys.\ J.\ {\bf 382},
 652 (1991).

\bibitem{Moskalenko:1993ke}
  I.~V.~Moskalenko and S.~Karakula,
   ``Very high-energy neutrinos from the sun,''
  J.\ Phys.\ G {\bf 19}, 1399 (1993).

\bibitem{Ingelman:1996mj}
G.~Ingelman and M.~Thunman, ``High Energy Neutrino Production by
Cosmic Ray Interactions in the Sun,'' Phys.\ Rev.\ D {\bf 54},
4385 (1996) [hep-ph/9604288].

\bibitem{Hettlage:1999zr}
C.~Hettlage, K.~Mannheim and J.~G.~Learned, ``The sun as a high
energy neutrino source,'' Astropart.\ Phys.\  {\bf 13}, 45 (2000)
[astro-ph/9910208].

\bibitem{Learned:2000sw}
  J.~G.~Learned and K.~Mannheim,
``High-energy neutrino astrophysics,''
  Ann.\ Rev.\ Nucl.\ Part.\ Sci.\  {\bf 50},  679 (2000).

  \bibitem{Halzen:2006mq}
  F.~Halzen,
  ``Astroparticle physics with high energy neutrinos: From AMANDA to IceCube,''
  astro-ph/0602132.


\bibitem{Srednicki:1986vj}
  M.~Srednicki, K.~A.~Olive and J.~Silk,
  `` High-energy neutrinos from the Sun and
 cold dark matter,''
  Nucl.\ Phys.\ B {\bf 279}, 804 (1987).

\bibitem{Ellis:1992df}
  J.~R.~Ellis, R.~A.~Flores and S.~S.~Masood,
  ``MSW effect on high-energy solar neutrinos from relic annihilation,''
  Phys.\ Lett.\ B {\bf 294}, 229 (1992).


  \bibitem{Bergstrom:1998xh}
  L.~Bergstrom, J.~Edsjo and P.~Gondolo,
  ``Indirect detection of dark matter in km-size neutrino telescopes,''
  Phys.\ Rev.\ D {\bf 58}, 103519 (1998)
  [hep-ph/9806293].

    \bibitem{deGouvea:2000un}
  A.~de Gouvea,
  ``The oscillation probability of GeV solar neutrinos of all active
  species,''
  Phys.\ Rev.\ D {\bf 63}, 093003 (2001)
  [hep-ph/0006157].

\bibitem{Crotty:2002mv}
  P.~Crotty,
  ``High-energy neutrino fluxes from supermassive dark matter,''
  Phys.\ Rev.\ D {\bf 66}, 063504 (2002)
  [hep-ph/0205116].

\bibitem{Cirelli:2005gh}
  M.~Cirelli {\it et al.},
  ``Spectra of neutrinos from dark matter annihilations,''
  Nucl.\ Phys.\ B {\bf 727}, 99 (2005)
  [hep-ph/0506298].

\bibitem{Halzen:2005ar}
  F.~Halzen and D.~Hooper,
  ``Prospects for detecting dark matter with neutrino telescopes in light of
  recent results from direct detection experiments,''
  Phys.\ Rev.\ D {\bf 73}, 123507 (2006)
  [hep-ph/0510048].


\bibitem{Sirignano}
  C.~Sirignano for the OPERA Collaboration,
  ``The CNGS project and  OPERA experiment at LNGS,'' talk at \emph{Neutrino 2006},
  22nd International Conference on Neutrino Physics and Astrophysics, available at:
   \texttt{neutrinosantafe06.com}.
   
\bibitem{Sjostrand:1995iq}
  T.~Sjostrand,
  ``PYTHIA 5.7 and JETSET 7.4: Physics and manual,''
  Comput.\ Phys.\ Commun.\ {\bf 82}, 74 (1994) [hep-ph/9508391].

\bibitem{cross}
  R.~Gandhi, C.~Quigg, M.~H.~Reno and I.~Sarcevic,
  ``Neutrino interactions at ultrahigh energies,''
  Phys.\ Rev.\ D {\bf 58}, 093009 (1998)
  [hep-ph/9807264];
  S.~Dutta, R.~Gandhi and B.~Mukhopadhyaya,
  ``Tau-neutrino appearance searches using neutrino beams from muon storage  rings,''
  Eur.\ Phys.\ J.\ C {\bf 18}, 405 (2000)
  [hep-ph/9905475].

\bibitem{Naumov:2001ci}
  V.~A.~Naumov,
  ``High-energy neutrino oscillations in absorbing matter,''
  Phys.\ Lett.\ B {\bf 529}, 199 (2002)
  [hep-ph/0112249].


\bibitem{Groom:2001kq}
  D.~E.~Groom, N.~V.~Mokhov and S.~I.~Striganov,
   ``Muon stopping power and range tables 10-MeV to 100-TeV,''
  Atom.\ Data Nucl.\ Data Tabl.\  {\bf 78}, 183 (2001).

\bibitem{Fogli2006}
 G.~L.~Fogli {\it et al.},
   ``Observables sensitive to absolute neutrino masses: A reappraisal after
  WMAP-3y and first MINOS results,''
  [hep-ph/0608060]; see also:
  G.~L.~Fogli, E.~Lisi, A.~Marrone and A.~Palazzo,
  ``Global analysis of three-flavor neutrino masses and mixings,''
  Prog.\ Part.\ Nucl.\ Phys.\  {\bf 57}, 742 (2006)
  [hep-ph/0506083].

\bibitem{Matt}  L.~Wolfenstein,
                ``Neutrino Oscillations In Matter,''
                Phys.\ Rev.\ D {\bf 17}, 2369 (1978);
                S. P.~Mikheev and A. Yu.\ Smirnov,
                ``Resonance Enhancement Of Oscillations In Matter And Solar Neutrino
                Spectroscopy,''
                Yad.\ Fiz.\ {\bf 42}, 1441 (1985)
                [Sov.\ J.\ Nucl.\ Phys.\ {\bf 42}, 913 (1985)].

\bibitem{KuoPa}
  T.~K.~Kuo and J.~T.~Pantaleone,
  ``Neutrino Oscillations In Matter,''
  Rev.\ Mod.\ Phys.\  {\bf 61}, 937 (1989).


  \bibitem{Bahcall:2004pz}
  J.~N.~Bahcall, A.~M.~Serenelli and S.~Basu,
  ``New solar opacities, abundances, helioseismology, and neutrino fluxes,''
  Astrophys.\ J.\  {\bf 621}, L85 (2005)
  [astro-ph/0412440].

\bibitem{Beacom:2003nh}
  J.~F.~Beacom {\it et al.}, 
  ``Measuring flavor ratios of high-energy astrophysical neutrinos,''
  Phys.\ Rev.\ D {\bf 68}, 093005 (2003)
  [Erratum-ibid.\ D {\bf 72}, 019901 (2005)]
  [hep-ph/0307025].

\bibitem{Serpico:2005bs}
  P.~D.~Serpico,
   ``Probing the 2-3 leptonic mixing at high-energy neutrino telescopes,''
  Phys.\ Rev.\ D {\bf 73}, 047301 (2006)
  [hep-ph/0511313].

\bibitem{Serpico:2005sz}
  P.~D.~Serpico and M.~Kachelrie\ss,
   ``Measuring the 13-mixing angle and the CP phase with neutrino  telescopes,''
  Phys.\ Rev.\ Lett.\  {\bf 94}, 211102 (2005)
  [hep-ph/0502088].

\bibitem{Kachelriess:2006fi}
  M.~Kachelriess and R.~Tomas,
``High energy neutrino yields from astrophysical sources. I: Weakly magnetized sources,''
  Phys.\ Rev.\ D {\bf 74}, 063009 (2006)
  [astro-ph/0606406].


  \bibitem{Athar:2000yw}
  H.~Athar, M.~Jezabek and O.~Yasuda,
  ``Effects of neutrino mixing on high-energy cosmic neutrino flux,''
  Phys.\ Rev.\ D {\bf 62}, 103007 (2000)
  [hep-ph/0005104].

\bibitem{quasivac}
  G.~L.~Fogli, E.~Lisi, D.~Montanino and A.~Palazzo,
  ``Quasi-vacuum solar neutrino oscillations,''
  Phys.\ Rev.\ D {\bf 62}, 113004 (2000)
  [hep-ph/0005261];  E.~Lisi {\it et al.}, 
  ``Analytical description of quasivacuum oscillations of solar neutrinos,''
  Phys.\ Rev.\ D {\bf 63}, 093002 (2001)
  [hep-ph/0011306].


\bibitem{Akhmedov:2005zh}
  E.~K.~Akhmedov, M.~Maltoni and A.~Y.~Smirnov,
  ``Describing oscillations of high energy neutrinos in matter precisely,''
  Phys.\ Rev.\ Lett.\  {\bf 95}, 211801 (2005) [hep-ph/0506064].


\bibitem{Peres:1999yi}
  O.~L.~G.~Peres and A.~Yu.~Smirnov,
  ``Testing the solar neutrino conversion with atmospheric neutrinos,''
  Phys.\ Lett.\ B {\bf 456}, 204 (1999)
  [hep-ph/9902312].


\bibitem{EGRETmoon}
D.~J.\ Thompson, D.~L.\ Bertsch,  D.~J.\ Morris and R.\ Mukherjee,
``Energetic gamma ray experiment telescope high-energy gamma ray
observations of the Moon and quiet Sun,''
Journal of Geophys.\ Res.\ {\bf 102} (A7), 14735 (1997).



\bibitem{GLAST}
See the website: \texttt{www-glast.slac.stanford.edu}

\bibitem{Moskalenko:2006ta}
  I.~V.~Moskalenko, T.~A.~Porter and S.~W.~Digel,
``Inverse Compton scattering on solar photons, heliospheric modulation, and
neutrino astrophysics,''  Astrophys.\ J.\ Letters, in press  [astro-ph/0607521].

\bibitem{Ambrosio:2003mz}
  M.~Ambrosio {\it et al.}  [MACRO Collaboration],
   ``Moon and sun shadowing effect in the MACRO detector,''
  Astropart.\ Phys.\  {\bf 20}, 145 (2003)
  [astro-ph/0302586].

  \bibitem{Barenboim}
G.~Barenboim and C.~Quigg,
``Neutrino observatories can characterize cosmic sources and neutrino
properties,''
Phys.\ Rev.\ D {\bf 67}, 073024 (2003) [hep-ph/0301220].



\bibitem{decay}
J.~F.~Beacom {\it et al.}, 
``Decay of high-energy astrophysical neutrinos,''
Phys.\ Rev.\ Lett.\ {\bf 90}, 181301 (2003)  [hep-ph/0211305];
J.~F.~Beacom {\it et al.}, 
``Pseudo-Dirac neutrinos, a challenge for neutrino telescopes,''
Phys.\ Rev.\ Lett.\  {\bf 92}, 011101 (2004) [hep-ph/0307151];
  D.~Hooper, D.~Morgan and E.~Winstanley,
  ``Lorentz and CPT invariance violation in high-energy neutrinos,''
  Phys.\ Rev.\ D {\bf 72}, 065009 (2005)
  [hep-ph/0506091].

\bibitem{Stanev:2004ys}
  T.~Stanev,
  {\em High Energy Cosmic Rays\/} (Springer-Praxis, Berlin, 2004).


\bibitem{Gu97}
Y.~Gu \emph{et al.},  ``A Stochastic Model of the Solar Atmosphere,''
 Astrophys.\ J. {\bf 484}, 960 (1997).



\end{thebibliography}
\end{document}